\documentclass[conference]{IEEEtran}
\IEEEoverridecommandlockouts
\usepackage{cite}
\usepackage{amsmath,amssymb,amsfonts}
\usepackage{pifont}
\usepackage{algorithmic}
\usepackage{graphicx}
\usepackage{url}
\usepackage{booktabs}
\usepackage{cprotect}
\usepackage{float}
\usepackage{multirow}
\usepackage{textcomp}
\usepackage{makecell}
\usepackage[table, dvipsnames]{xcolor}
\usepackage{array}
\usepackage{threeparttable}
\def\BibTeX{{\rm B\kern-.05em{\sc i\kern-.025em b}\kern-.08em
    T\kern-.1667em\lower.7ex\hbox{E}\kern-.125emX}}
\definecolor{lightgrayrow}{HTML}{EFEFEF}

\usepackage{tikz}
\usepackage{circledsteps}

\definecolor{circBG}{RGB}{186,90,84} %

\DeclareRobustCommand{\circled}[1]{%
  \tikz[baseline=(char.base)]{
    \node[circle, fill=circBG, inner sep=0pt] (char)
    {\color{white}\strut #1};
  }%
}

\usepackage{titlesec}

\titlespacing{\section}{0pt}{1.0ex plus 0.2ex minus 0.2ex}{0.7ex plus 0.2ex}
\titlespacing{\subsection}{0pt}{0.8ex plus 0.2ex minus 0.2ex}{0.5ex plus 0.2ex}
\titlespacing{\subsubsection}{0pt}{0.6ex plus 0.2ex minus 0.2ex}{0.4ex plus 0.2ex}

\begin{document}

\title{Demystifying NVSHMEM: A System-Level Analysis on Symmetric Memory and Device-Initiated Operations in GPU Communication}

\author{
\IEEEauthorblockN{
Yijun Ma$^1$\IEEEauthorrefmark{1},
Siyuan Shen$^1$\IEEEauthorrefmark{1},
Tiancheng Chen$^1$,\\
Akhil Langer$^2$,
Jiri Kraus$^2$,
Benjamin Glick$^2$,
Craig Belusar$^2$,
Jeff Hammond$^2$,
Torsten Hoefler$^1$
}

\IEEEauthorblockA{$^1$ETH Z{\"u}rich, Switzerland, \{siyuan.shen, tiancheng.chen, torsten.hoefler\}@inf.ethz.ch, yijunma@student.ethz.ch}

\IEEEauthorblockA{$^2$NVIDIA Corporation, \{alanger, jkraus, benjaming, cbelusar, jeffpapers\}@nvidia.com}
\IEEEauthorblockA{* These authors contributed equally to this work}
}

\maketitle

\begin{abstract}

NVSHMEM is NVIDIA's OpenSHMEM-based PGAS communication library for GPU clusters, enabling GPU-initiated, one-sided communication through symmetric memory. Despite its growing adoption, a system-level understanding of its design and behavior remains scattered across documentation, source code, and application experience. This paper presents a concise study of NVSHMEM's programming model, implementation, and performance characteristics, focusing on symmetric memory, one-sided operations, and device-side collectives. We also examine DeepEP as a case study of NVSHMEM in performance-critical sparse deep learning workloads. Our analysis shows that NVSHMEM pioneered a device-side symmetric-memory programming model that enables fine-grained GPU-driven communication and is important for approaching the hardware performance limit. Overall, this work defines NVSHMEM's role as a systems building block, highlights its design tradeoffs, and identifies opportunities for improving GPU communication runtimes.
\end{abstract}


\section{Introduction}

The rise of distributed deep learning and GPU-accelerated scientific computing has made GPU communication central. Distributed deep learning relies on data, tensor, and pipeline parallelism to scale training and inference across many GPUs and often multiple nodes~\cite{bennun2020demystifyingparallel}. HPC applications likewise use distributed decompositions to scale simulations beyond a single device~\cite{valdez2013towardsaccelearting, frances2015multigpu, punke2026multigpufastfouriertransforms}. In both settings, performance depends on efficient data exchange over PCIe, NVLink, and InfiniBand (IB). Communication libraries must therefore provide high bandwidth, low latency, and abstractions that fit GPU execution. The NVIDIA Collective Communication Library (NCCL)~\cite{hu2026demystifyingnccl} has become the dominant solution, providing optimized collectives and serving as the recommended backend for CUDA-based distributed ML frameworks such as PyTorch~\cite{paszke2019pytorch}, Megatron-LM~\cite{shoeybi2020megatronlm}, and vLLM~\cite{kwon2023vllm}.

Originally, NCCL was host-driven: the CPU enqueues collectives, and the library selects algorithms and schedules. While effective for bulk-synchronous collectives, this model is less suitable for fine-grained point-to-point communication. Such patterns arise when communication is data-dependent, restricted to subsets of threads, or sensitive to CPU coordination latency. Examples include stencil halo exchange, irregular graph communication, sparse and expert-parallel workloads, and custom kernels that tightly couple computation and data movement. 

NVSHMEM~\cite{nvshmem-api-documentation, maruyama2020nvshmem} addressed this gap through a complementary programming model. It exposes a partitioned global address space (PGAS) directly to GPU code, enabling CUDA kernels to access symmetric memory on remote processing elements via one-sided put/get operations, atomics, and explicit synchronization primitives. Rather than replacing NCCL, NVSHMEM enabled communication patterns that are difficult to express with host-launched collectives, including irregular exchanges and fine-grained overlap.

As NVSHMEM proved useful across several workloads, NCCL incorporated related ideas through its new device API, adding device-initiated operations and symmetric memory support~\cite{hamidouche2025gin}. These features build on NVSHMEM concepts and reflect lessons from AI workloads, but the two designs target different abstractions. NVSHMEM provides a flat remote-memory view
while NCCL makes the distinction between scale-up and scale-out domains explicit. NCCL’s device API also addresses a key limitation of NVSHMEM: the lack of hierarchical composition through communicators. Although NVSHMEM supports teams, its OpenSHMEM roots still expose a single symmetric heap shared across all GPUs in an NVSHMEM instance.

Despite NVSHMEM’s growing importance, many key behaviors are not apparent from the API alone. How does NVSHMEM realize symmetric memory on modern GPUs? Which communication paths avoid CPU involvement, and when are they used? What mechanisms underlie its collectives? We answer these questions through a source-level analysis of NVSHMEM, tracing its design from the programming model to runtime internals and transport mechanisms. Comparisons with NCCL clarify design differences rather than provide an official ranking.

Our implementation analysis targets NVSHMEM version 3.3.9; the microbenchmarks evaluate version 3.8.0 on B200 GPUs. Although releases may change specific features, core components such as symmetric memory and communication paths are fundamental to NVSHMEM's design and likely to remain largely unchanged. We therefore expect the main insights to remain relevant.

\section{Background}

\subsection{Essential Concepts}

We first introduce the core concepts shaping NVSHMEM: limitations of host-initiated communication, motivation for device-initiated communication, and the PGAS/OpenSHMEM model based on symmetric memory and one-sided operations.

\textbf{NVIDIA Collective Communication Library (NCCL)} is the standard communication library for multi-GPU systems, but its conventional interface remains largely \textbf{host-initiated}~\cite{hu2026demystifyingnccl}. Communicator setup and collective launches are performed from the CPU, which schedules NCCL kernels onto CUDA streams. Although optimized for regular collectives, this model keeps the host in the communication control path and can add coordination overhead.

This limitation motivates \textbf{device-initiated communication}, where GPU code launches communication and synchronization directly instead of through a host-managed control path. This reduces CPU involvement and enables finer-grained overlap between communication and computation. These benefits matter for latency-sensitive and strongly scaled GPU workloads, motivating interfaces such as NVSHMEM that support GPU-side communication primitives and GPU-initiated network operations that bypass CPU proxy threads~\cite{ibgda, nvshmem-api-documentation, potluri2017gpucentric}.

\textbf{Peer-to-peer (P2P) memory access} means direct access from one GPU to another GPU's memory, distinct from point-to-point messaging or peer-to-peer network architectures. This enables low-overhead remote memory access within a \textbf{scale-up} domain, usually node-level NVLink/NVSwitch but potentially multi-node NVLink (MNNVL). \textbf{GPUDirect RDMA (GDRDMA)}~\cite{nvidia_gpudirect_rdma} extends direct GPU memory access to \textbf{scale-out} domains by allowing an RDMA-capable NIC to read and write GPU memory. \textbf{GPUDirect Async Kernel-Initiated (GDA-KI)}~\cite{nvidia_gdaki, hamidouche2025gin} further lets GPU kernels initiate and control network operations without CPU intervention. \textbf{InfiniBand GPUDirect Async (IBGDA)}~\cite{ibgda} is the InfiniBand-specific realization of GDA-KI.

The \textbf{Partitioned Global Address Space (PGAS)} model exposes a logically shared address space partitioned across processing elements (PEs)~\cite{calin2013theory-pgas}. It underlies OpenSHMEM~\cite{chapman2010openshmem} and is adapted by NVSHMEM for GPU systems. It is also related to MPI-3 RMA~\cite{hoefler2015mpi3rma}, where a process can access memory exposed by another process without an explicit matching receive. The main difference is the interface and runtime model: OpenSHMEM and NVSHMEM encode operation properties such as data type, transfer size, and operation kind directly into API names, while MPI RMA passes them as arguments. This SHMEM-style interface can reduce dispatch and argument-processing overhead, which helps fine-grained communication.

In NVSHMEM, a PE is an OS process mapped to one GPU; since version 2.4.1, NVSHMEM also provides limited support for multiple PEs per GPU. Communication uses \textbf{symmetric memory}: a \textbf{symmetric heap} on each PE and \textbf{symmetric objects} with the same type, size, and layout across PEs. Remote PEs access these objects through one-sided operations such as \texttt{put} and \texttt{get}, where the initiator specifies the source, destination, and location without explicit receives from the target PE. This enables asynchronous progress and, often, direct placement into the destination’s symmetric memory buffer.

\subsection{GPU Communication Libraries}

To understand NVSHMEM’s role in the broader high-performance computing (HPC) landscape, we compare it with existing GPU communication frameworks and libraries.

\textbf{CUDA-aware MPI}~\cite{kraus2025cudaawarempi} extends the standard MPI programming model~\cite{mpi-standard} by allowing MPI routines to operate directly on GPU-resident buffers, avoiding explicit application-level staging through host memory. This makes it an effective path for porting existing distributed-memory HPC applications to GPU clusters while preserving familiar rank-based, two-sided communication semantics \cite{mpi-standard}. However, CUDA-aware MPI remains fundamentally host initiated.

\textbf{NCCL}, discussed previously, is the main alternative to NVSHMEM for GPU communication. It is optimized for dense and regular collective communication~\cite{hu2026demystifyingnccl}. This makes it the natural baseline for distributed deep learning and other bulk-synchronous workloads. However, its conventional interface was host-centric. Recent NCCL Device API work reduces this gap by introducing GPU-initiated communication, but the use of NCCL in the AI ecosystem remains collective-centric.  NVSHMEM was designed around fine-grained one-sided operation and currently possesses functionality not found in NCCL's GPU-Initiated Networking (GIN) interface~\cite{hamidouche2025gin}.

\textbf{UPC++}~\cite{upcxx} and \textbf{GASNet-EX}~\cite{BonacheaHargrove2018_GASNetEX} provide portable PGAS-style communication and share with NVSHMEM the idea of one-sided access to a partitioned address space. Their portability and C++ integration make them attractive for many host-driven HPC applications. The main distinction, however, lies in where communication is initiated. While UPC++ and GASNet-EX can support GPU memory through device-aware transports, the control path is still typically managed by host code. NVSHMEM, by contrast, exposes device-callable operations that allow CUDA threads, warps, or thread blocks to initiate communication directly from the GPU.

\textbf{rocSHMEM}~\cite{rocshmem} and \textbf{Intel SHMEM}~\cite{brooks2024intelshmem} bring OpenSHMEM-style, GPU-oriented one-sided communication to AMD ROCm and Intel oneAPI/SYCL systems, respectively. Together, they show that device-initiated PGAS communication is becoming a broader vendor trend rather than a feature unique to NVIDIA. However, each implementation is necessarily shaped by its own compiler stack, memory model, interconnects, and transport mechanisms.  GICC~\cite{shan2026gicc}  implements GPU-initiated communication for the Slingshot network using triggered operations.

\section{NVSHMEM Overview}

Before analyzing NVSHMEM's internals, we first summarize the main concepts that shape how it is used in practice.

\subsection{Teams}

Teams are NVSHMEM’s mechanism for defining subsets of processing elements (PEs) that participate in communication. Unlike MPI or NCCL communicators, teams do not contain most of the communication-related runtime state. Instead, they are lightweight handles that identify PE groups and may encode topology information. The default team is \texttt{NVSHMEM\_TEAM\_WORLD}, which contains all PEs and is implicitly used whenever a team argument is omitted.

Teams carry both semantic and runtime constraints. A collective on a team involves exactly the PEs in that team, and team-relative operations must translate PE indices correctly. Arbitrary teams can also be constructed at runtime from existing PEs. However, a team cannot safely be used by multiple concurrent collective invocations from the same PE because the runtime associates a single set of internal resources, such as synchronization state and scratch storage, with each team.

\subsection{API and Dual-Interface Design}

This dual-interface design is one of NVSHMEM’s key strengths, but it also makes the library more complex than communication libraries built around a single control path. This complexity is compounded by the fact that the split is not uniform across the API: some components remain fundamentally host-managed, whereas the core communication path is exposed on both the CPU and the GPU. To clarify this structure, we group the APIs into subsets and summarize their purpose and availability in Table~\ref{tab:api-groups} in the Appendix.

In addition, the naming convention distinguishes the OpenSHMEM interface from NVSHMEM-specific extensions: \texttt{nvshmem\_} generally denotes the standard API, whereas \texttt{nvshmemx\_} is used for extensions such as stream-ordered operations and threadgroup-scoped variants. Internal helper functions often use the \texttt{nvshmemi\_} prefix, where the \texttt{i} denotes internal routines that are not part of the public API.

Within the device-side APIs, NVSHMEM distinguishes operations by execution scope. The default APIs are thread-scoped, where a single CUDA thread issues the operation. Many \texttt{nvshmemx\_} extensions also provide \texttt{\_warp} and \texttt{\_block} variants, where the calling warp or block collectively provides the resources needed to execute one operation.

We present an overview of the functionality of each API group as follows. For a complete description, we refer the reader to the official NVSHMEM documentation~\cite{nvshmem-api-documentation}.

\subsubsection{Setup / exit}

This group brings NVSHMEM up and tears it down. It covers initialization, bootstrap through MPI or unique IDs, and finalization, and is primarily responsible for creating the runtime state required by all later operations. In the source, this group is mostly host-side, with \texttt{nvshmem\_global\_exit} as the main exception that is also exposed on the device.

\noindent\subsubsection{Memory management}

The memory-management APIs allocate and manage symmetric objects in the symmetric heap. They include routines such as \texttt{nvshmem\_malloc}, \texttt{nvshmem\_align}, and \texttt{nvshmem\_free}, together with buffer-registration functionality for selected local buffers. These routines remain host-driven because they manipulate global runtime state and collective memory layout.

\subsubsection{Team management}

Team management APIs define communication subgroups and provide the rank translation needed. Functions such as \texttt{nvshmem\_team\_my\_pe} are available on both host and device, while team creation and destruction remain host-side.

\subsubsection{One-sided remote memory access (RMA)}

The RMA group is the core one-sided communication interface of NVSHMEM. It includes typed and untyped \texttt{put}/\texttt{get} operations, scalar \texttt{p}/\texttt{g}, strided \texttt{iput}/\texttt{iget}, and nonblocking variants. The API appears in several generalized naming forms: typed routines follow patterns such as \texttt{nvshmem\_<TYPENAME>\_\{put,get\}}, size-specialized routines follow forms such as \texttt{nvshmem\_\{put,get\}<SIZE>}, and untyped byte-oriented routines use names such as \texttt{nvshmem\_\{putmem,getmem\}}. Scalar \texttt{p}/\texttt{g} operations cover single-element access, while \texttt{iput}/\texttt{iget} provide shorthand for strided communication patterns. This group is fully dual-interface and is also extended with device-side warp- and block-scoped variants and host-side on-stream forms.

\subsubsection{Atomics}

Atomic APIs provide remote read-modify-write operations on symmetric data, including operations such as fetch and add. Their purpose is to support fine-grained remote coordination on individual memory locations. The API follows generalized naming forms such as \texttt{nvshmem\_<TYPENAME>\_}\texttt{atomic\_<op>} for non-fetching operations and \texttt{nvshmem\_<TYPENAME>\_} \texttt{atomic\_fetch\_<op>} for fetching variants, along with special forms such as \texttt{compare\_swap}, \texttt{swap}, and \texttt{fetch}.

\subsubsection{Memory ordering}

The memory-ordering interface is centered around a small number of key routines. \texttt{nvshmem\_fence} enforces the ordering of previously issued communication operations to the same destination PE but does not, by itself, guarantee their completion. \texttt{nvshmem\_quiet} is stronger: it waits until previously issued NVSHMEM communication operations from the calling context have completed and become visible at the destination. On the host side, NVSHMEM also provides \texttt{nvshmemx\_quiet\_on\_stream}, which places this completion semantics into CUDA stream order.

\subsubsection{Synchronization}

The synchronization group lets a PE wait for or test remote progress based on the value of a symmetric object or signal variable. This includes the \texttt{wait} and \texttt{test} families as well as signal-based waiting operations. This group is less uniformly dual-interface: the ordinary \texttt{wait}/\texttt{test} routines are primarily device-side, while the host mainly accesses this functionality through on-stream variants.

\subsubsection{Collectives}

The NVSHMEM collective API group provides team-wide communication and synchronization operations, including Barrier, Sync, Broadcast, AlltoAll, FCollect, Reduce, and ReduceScatter. These APIs are available on both host and device. FCollect is the counterpart of MPI AllGather, while NVSHMEM Reduce is the counterpart of MPI AllReduce. In addition, NVSHMEM memory-management and team-creation functions are also collective and internally invoke \verb|nvshmem_barrier_all|.

\section{Memory Management}

Having outlined the overall design of NVSHMEM, we now examine its internals in more detail. We begin with how symmetric memory is realized and managed, as it is one of the core foundations enabling device-initiated communication.

\begin{figure}[t]
  \centering
  \includegraphics[width=\linewidth]{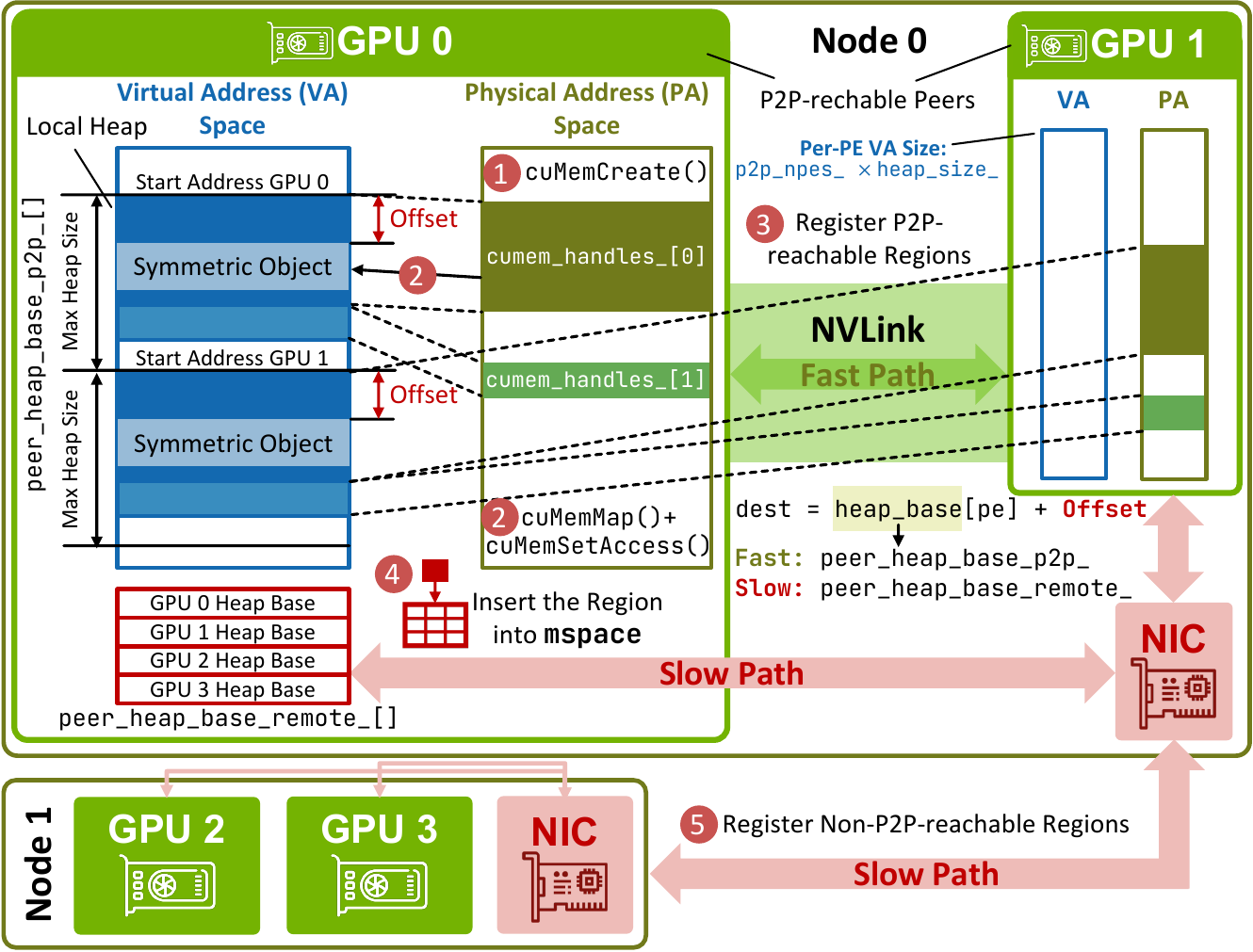}
  \vspace{-2em}
  \cprotect\caption{Virtual-memory-based symmetric-heap setup in NVSHMEM for two nodes, each with two GPUs and one NIC. The numbered markers \circled{1}--\circled{5} show the main steps taken when new physical memory is allocated.
  }
  \label{fig:memory-management}
  \vspace{-1.5em}
\end{figure}

NVSHMEM implements the symmetric heap on top of CUDA's low-level virtual memory management (VMM) API, and alternative backends using system memory or pinned device memory can be selected via environment options. With VMM enabled, NVSHMEM implements memory management scheme that cleanly separates \textbf{virtual address (VA) reservation} from \textbf{physical memory allocation}. The symmetric heap's VA range is reserved eagerly during initialization, while physical pages are committed on demand as the application allocates from the heap~\cite{perry2020introducingvmm, langer2022dynamic}. Figure~\ref{fig:memory-management} illustrates the key concepts discussed in this section.

\subsection{Initialization}

During \verb|nvshmem_init|, NVSHMEM creates a symmetric heap for each PE. It first chooses the allocation granularity and sets the per-PE heap size to the larger of \verb|MAX_MEMORY_| \verb|PER_GPU| and the internal runtime memory overhead, rounded up to the chosen granularity. Each PE then reserves a contiguous VA range of size \verb|p2p_npes_| \(\times\) \verb|heap_size_| using \verb|cuMemAddressReserve()|, where \verb|p2p_npes_| is the number of P2P-reachable GPUs. The first segment is used for the local heap, while the remaining equal-sized segments are used for peer mappings. NVSHMEM then creates a host-side \verb|mspace| allocator without committing physical memory. After transport initialization, PEs exchange their local heap base addresses so that remote memory access can be established.

The reserved VA range covers only P2P-mappable peers, so its maximum size scales with \verb|p2p_npes_|. This keeps the direct-mapping scheme practical for the scale-up domain. For non-P2P peers, NVSHMEM instead allgathers each PE’s \verb|heap_base_| into a separate table, \verb|peer_heap_base_| \verb|remote_|, which stores the actual base address of every remote heap. The slow path uses this table together with transport metadata and remote memory handles.

\subsection{Memory Allocation \& Registration}

On top of the reserved VA range, NVSHMEM manages the heap with a simple host-side allocator. It maintains three \verb|std::map| structures: two for free chunks indexed by start and end addresses, and one for allocated chunks indexed by the start address. This design makes coalescing straightforward since freeing a block allows the allocator to check both neighboring directions and merge adjacent free chunks when possible. Allocation is performed using a first-fit policy: the free list is scanned in address order, the first sufficiently large block is chosen, and it is either split or consumed entirely. All sizes are aligned to \verb|NVSHMEMI_MALLOC_ALIGNMENT|, and the address is returned to the user.

The allocator is deliberately simple and deterministic. It linearly scans the free list, and allocation is \(O(n)\) in the number of free chunks. This overhead is acceptable as allocations are typically executed during initialization. In contrast to allocators such as \texttt{jemalloc}, it avoids per-thread arenas, size classes, or other complex optimizations.

NVSHMEM allocates physical GPU memory for the symmetric heap only on demand, when an allocation request cannot be satisfied from already-backed VA space. When a symmetric allocation request triggers the growth of the heap, \verb|allocate_physical_memory_to_heap()| performs the following steps:

\noindent\circled{1} Create a physical memory handle with \verb|cuMemCreate()|.

\noindent\circled{2} Map the corresponding heap subrange with \verb|cuMemMap()| and install access permissions with \verb|cuMemSetAccess()| for the local and eligible peer GPUs.

\noindent\circled{3} Register the region with P2P transport.

\noindent\circled{4} Insert the region into the \verb|mspace| allocator.

\noindent\circled{5} Register the region with network transports for communication with non-P2P-reachable peers.

Heap growth is therefore incremental: NVSHMEM maps and registers the newly backed subrange rather than rebuilding the entire heap mapping. It then records the allocation metadata (i.e., offset, size, etc.), exchanges the required handles across PEs, refreshes device-resident metadata, and synchronizes with a barrier.

\subsection{Mapping and Remote Address Computation}

NVSHMEM reserves a VA range large enough to cover the symmetric heaps of all P2P-reachable GPUs and places each peer heap at a fixed offset within that range. This organization is central to the design, because it allows remote addresses to be derived from heap-relative offsets.

At a high level, it exports the local CUDA memory handle, exchanges these handles across PEs, and maps peer memory into the corresponding reserved virtual address segments. For non-P2P transports, the same region is additionally registered with the appropriate network interface.

Remote address computation follows the same symmetric-offset rule in both the fast and slow paths, the only difference is which per-PE base address is used and whether the resulting address can be dereferenced directly. In the P2P fast path, if \verb|heap_base_| is the base of the local symmetric heap and \verb|peer_heap_base_p2p_[remote_pe]| is the base at which the remote PE's heap is mapped in the local VA space, then for a local symmetric pointer \verb|dest_local| the corresponding remote address is
\begin{equation}
\small
\begin{aligned}
\texttt{dest\_remote}
= &\texttt{peer\_heap\_base\_p2p\_[remote\_pe]} \\
&+ (\texttt{dest\_local} - \texttt{heap\_base\_})
\end{aligned}
\label{eq:pointer-arithmetic}
\end{equation}
NVSHMEM, therefore, preserves the object's offset within the local heap and applies it to the remote heap base. For non-P2P-accessible GPUs, the same rule is applied using \verb|peer_heap_base_remote_[remote_pe]|, and the computed address is passed to the transport layer together with the required memory handle, registration metadata, or NIC key. The same addressing scheme thus applies in both paths, but only the P2P case gives the GPU a directly mapped peer virtual address. NVSHMEM's device-side fast path uses this mapping for SM-issued loads and stores from within the kernel, which is distinct from host-initiated CUDA peer copies, which may use asynchronous copy engines depending on the GPU and interconnect. Whenever the heap grows, NVSHMEM also refreshes device-resident metadata so that GPU kernels observe the updated layout correctly.

\section{One-sided Communication}

Having discussed how NVSHMEM manages memory, we now turn to another fundamental aspect of the library: one-sided communication. As noted in the previous section, NVSHMEM uses either a fast path or a slow path depending on whether the target peer is P2P-reachable. The following sections explain how data is moved between PEs in each case.

\subsection{Fast Path: Direct GPU Memory Access}

\begin{figure}[t]
  \centering
  \includegraphics[width=\linewidth]{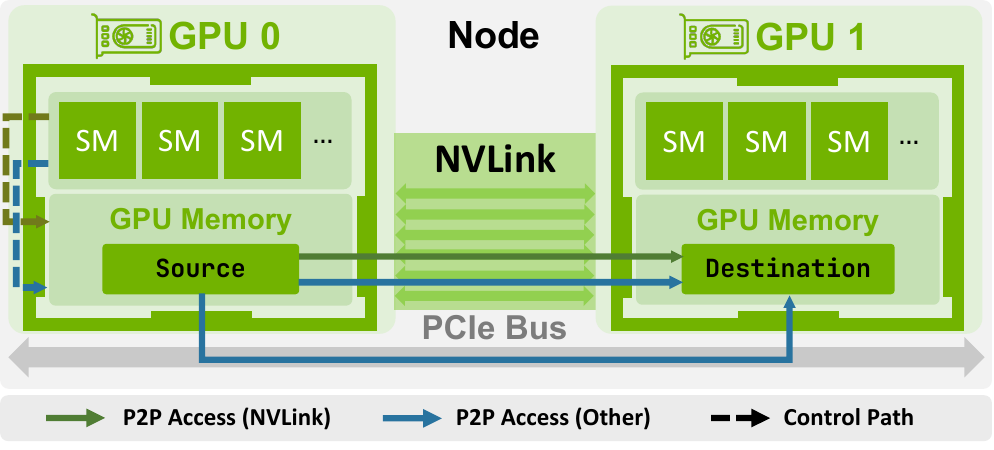}
  \cprotect\caption{Illustration of the device-side RMA fast path. Solid arrows denote data paths, while dashed arrows denote control paths, of the same color highlight the control path. Although the figure shows NVLink within a node, the same path applies beyond a single node when using MNNVL}
  \label{fig:fast-path}
  \vspace{-1.5em}
\end{figure}

\textbf{Fast-path eligibility.}
This decision is made by the host-side P2P transport logic. Figure~\ref{fig:fast-path} illustrates the two common cases that follow from this decision. For brevity, we focus on the most common case and omit less typical fallback paths, such as NVIDIA Management Library (NVML)-based handling. In this path, NVSHMEM verifies that the two PEs are on the same host, identifies the peer as a locally visible CUDA device, and uses \verb|cudaDeviceCanAccessPeer| to determine whether direct peer access is supported. If P2P access is available, it queries whether native GPU atomics can be enabled. Only then does NVSHMEM populate \verb|peer_heap_base_p2p_[pe]| for that PE.

\textbf{Device-side RMA}
Once the peer heap is directly mapped, device-side RMA reduces to ordinary memory access on the computed remote address from Equation~\ref{eq:pointer-arithmetic}. For scalar operations, helpers such as \verb|nvshmemi_p| and \verb|nvshmemi_g| first check whether the target PE has a mapped heap base and then issue a direct store or load on the resulting remote pointer. Similarly, bulk interfaces such as \verb|nvshmemi_put_threadgroup| perform threadgroup-wide \texttt{memcpy} operations to or from that mapped address.

\textbf{Host-side RMA}
Host-side RMA uses the same mapped address space, but the execution model is different since the host orchestrates CUDA copy operations. Public APIs such as \verb|nvshmem_put| and \verb|nvshmemx_put_on_stream| delegate to \verb|nvshmemi_prepare_and_post_rma|, which selects a mapped P2P path when the peer heap is locally accessible. In that case, NVSHMEM uses CUDA copy operations such as \verb|cudaMemcpyAsync| between local and remote device pointers. For device-local symmetric buffers, it may use specialized helpers such as \verb|nvshmemi_p2p_rma_optimized|, which select explicit copy kinds for stream-ordered and single-word operations.

\subsection{Slow Path: IBGDA and Proxy Execution}

\begin{figure}[t]
  \centering
  \includegraphics[width=\linewidth]{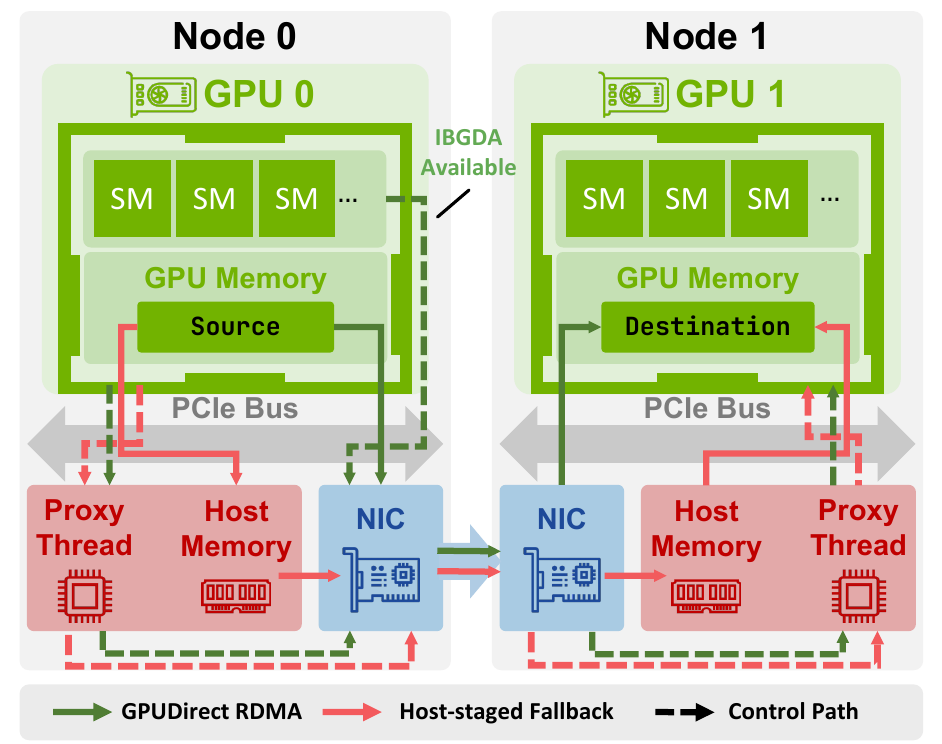}
  \cprotect\caption{Illustration of device-side RMA slow path. Solid arrows denote data movement, while dashed arrows of the same color highlight the control path.}
  \label{fig:slow-path}
  \vspace{-2em}
\end{figure}

The slow path is used when the target PE's symmetric heap is not directly P2P-mapped into the local GPU address space. This occurs for inter-node communication and for intra-node peers that are not CUDA P2P-accessible. In this case, NVSHMEM still preserves the same offset-based addressing scheme described earlier, but the computed remote address can no longer be dereferenced directly by the local GPU. Instead, the operation is carried out through a network transport, as illustrated in Figure~\ref{fig:slow-path}. Depending on system support, NVSHMEM either uses InfiniBand GPUDirect Async (IBGDA), if available, or a \textbf{host proxy} path, where a CPU thread executes the request on behalf of the GPU.

\textbf{Network transport and connection management.}
Slow-path RMA relies on a pluggable remote-transport layer rather than on a single network interface model. NVSHMEM can select among InfiniBand-oriented transports such as IBRC and IBDEVX, as well as higher-level backends such as UCX and libfabric. The InfiniBand-specific transports maintain PE-to-PE connectivity through Reliable Connection (RC) queue pairs (QPs), while IBGDA additionally supports Dynamic Connection (DC) queue pairs.

However, this RC-QP description does not apply universally. For example, on HPE Slingshot systems, NVSHMEM uses the libfabric transport with the \texttt{cxi} provider, where connectivity is established through libfabric endpoints. UCX similarly constructs its own PE-to-PE endpoints from exchanged worker addresses. Recent NVSHMEM versions also expose QP-specific APIs to users, allowing finer control over QP selection and usage when IB-specific paths are active~\cite{nvshmem-api-documentation}.

\begin{table*}[t]
\centering
\caption{Summary of collective algorithms in NVSHMEM (v3.3.9). $M$ is the message size, $N$ is the number of PEs, and $S=\lceil M / B_{\mathrm{seg}} \rceil$ denotes the number of scratch-limited segments in segmented AllReduce. Algorithms whose names contain \texttt{NVLS} rely on NVSwitch and NVLink SHARP to perform in-network operations.}
\label{tab:collective-algorithms}
\renewcommand{\arraystretch}{1.08}
\resizebox{1\textwidth}{!}{%
\begin{tabular}{cllll}
\toprule
\multicolumn{1}{c}{\textbf{Collective}} &
\multicolumn{1}{c}{\textbf{Algorithm}} &
\multicolumn{1}{c}{\textbf{LL / LL128}} &
\multicolumn{1}{c}{\textbf{Communication Volume}} &
\multicolumn{1}{c}{\textbf{Latency, \# Synchronizations}} \\
\midrule

\multirow{3}{*}{Broadcast}
& Bruteforce put-to-all & No & $O(MN)$ & $O(1)$ \\
& $k$-ary flat tree & LL only & $O(MN)$ & $O(\log_k N)$ \\
& Topology-aware hierarchical tree & LL only & $O(MN)$ & $O(\log_k N_{\mathrm{remote}} + \log_k N_{\mathrm{intra}})$ \\

\midrule
AlltoAll
& P2P / general all-push & No & $O(MN^2)$ & $O(1)$ \\

\midrule
\multirow{2}{*}{\makecell{FCollect \\ (AllGather)}}
& NVLS one-shot & LL only & $O(MN^2)$ & $O(1)$ \\
& Generic all-push & Both & $O(MN^2)$ & $O(1)$ \\

\midrule
\multirow{5}{*}{\makecell{Reduce \\ (AllReduce)}}
& NVLS one-shot & No & $O(MN^2)$ & $O(1)$ \\
& NVLS two-shot & No & $O(MN)$ & $O(1)$ \\
& $k$-ary recursive exchange & No & $O(M N \times k \log_k N)$ & $O(\log_k N)$ \\
& Hierarchical fcollect & Inherited from FCollect & $O(MN^2)$ & $O(1)$ \\
& (Segmented) linear AllReduce & No & $O(MN^2)$ & Direct LD/ST: $O(1)$; Segmented: $O(SN)$ \\

\midrule
\multirow{2}{*}{ReduceScatter}
& NVLS one-shot & No & $O(MN^2)$ & $O(1)$ \\
& Generic all-push & No & $O(MN^2)$ & $O(1)$ \\

\bottomrule
\end{tabular}%
}
\vspace{-2em}
\end{table*}

\textbf{Device-side RMA}
On the device side, scalar put/get operations use helpers such as \verb|nvshmemi_transfer_rma_p| and \verb|nvshmemi_transfer_|\verb|rma_g|; bulk operations use \verb|nvshmemi_transfer|\verb|_rma| and \verb|nvshmemi_| \verb|transfer_rma_nbi|; and related operations such as signals and atomics are handled through specialized helpers such as \verb|nvshmemi_transfer_put_signal| and \verb|nvshmemi_transfer_amo_*|.

Each operation first checks whether IBGDA is available on the device. If so, the request is offloaded to routines such as \verb|nvshmemi_ibgda_rma_*|, which construct and post RDMA work directly from GPU code to the NIC. Otherwise, the operation is encoded as a work request and written into a proxy buffer in host-pinned memory. A dedicated host proxy thread then consumes this descriptor and executes the corresponding operation via functions such as \verb|nvshmemi_proxy_rma_p| and \verb|nvshmemi_proxy_rma_g|. Thus, the slow path remains GPU-initiated at the API level, but completion is ultimately driven either by the NIC through IBGDA or by the host proxy through the descriptor queue.

\textbf{Host-side RMA}
Host-side RMA APIs, such as \verb|nvshmem_put|, delegate to the common helper \verb|nvshmemi_| \verb|prepare_and_post_rma|, which selects the appropriate transport. On the slow path, in short, the operation is treated as a remote network transfer.

In the off-stream case, NVSHMEM directly selects a remote transport with RMA capability, such as IBRC or libfabric, constructs the corresponding local and remote memory descriptors, and posts the operation through the transport's host-side RMA entry point. Bulk transfers are handled through \verb|nvshmemi_process_multisend_rma|, which posts the required sequence of RDMA operations. In the on-stream case, the request is instead encoded into the proxy command structure and launched on the specified CUDA stream through \verb|nvshmemi_proxy_rma_launcher|, after which the host proxy thread drives the actual network operations..

The host-side slow path has several remote-transport limitations: remote strided RMA and host-side \verb|g| operations are unsupported, and \verb|put_signal| is available only through the on-stream proxy path. These limitations reflect that the current remote transport and proxy mechanisms are primarily optimized for contiguous transfers and device-side signaling.

\section{Collective Communication}

Having established the mechanisms for symmetric memory and one-sided communication, we next examine NVSHMEM’s collective operations. These operations are built on top of the lower-level components discussed above.

NVSHMEM implements a variety of collective algorithms to target different message sizes and system configurations. Due to space constraints, we do not discuss each algorithm in detail. Instead, Table~\ref{tab:collective-algorithms} summarizes the main designs and, based on our analysis, reports their communication volume and latency, measured by the number of synchronization steps. In the remainder of this section, we therefore focus on the core concepts that underlie NVSHMEM’s collective implementation rather than on each algorithm individually.

\subsection{{\texttt{pSync} Buffer}}
Each team in NVSHMEM owns a \verb|pSync| buffer, short for \textbf{persistent synchronization buffer}, which is a symmetric memory region that is used to coordinate collective operations, and in some cases point-to-point operations, across PEs. NVSHMEM lays out \verb|pSync| regions in a strided manner across teams so that synchronization states belonging to different teams do not fall on the same cache lines. Within each team's \verb|pSync| region, different collective operations are assigned fixed subregions using hardcoded offsets. In addition, some operations are double-buffered so that consecutive invocations can alternate between buffers and avoid requiring an additional barrier between uses.

\subsection{Low Latency (LL) \& LL128 Protocols}

To reduce the cost of explicit synchronization after data transfer, which can become a major bottleneck for small messages, NVSHMEM implements Low Latency (LL) and LL128 protocols, following a similar design as NCCL~\cite{hu2026demystifyingnccl}. The central idea of these protocols is to couple data movement with lightweight arrival notification.

In the LL protocol, each data unit is paired with synchronization flags: the sender packs two data elements and two flags into a single 16-byte atomic write, and the receiver polls the flags to determine when the data is ready. LL128 uses the same idea with a larger transfer unit, grouping 120 bytes of data with an 8-byte flag into 128 bytes to improve bandwidth utilization. However, LL128 is only safe over NVLink because it relies on 128-byte atomic stores, which are not generally guaranteed on interconnects such as PCIe~\cite{hu2026demystifyingnccl}.

The LL and LL128 protocols require additional \texttt{pSync} storage, since the reception buffer must accommodate both payload data and the flags. As not all algorithms support these protocols, Table~\ref{tab:collective-algorithms} highlights the ones that do.

\subsection{Algorithm Selection}

In general, NVSHMEM selects collective algorithms at runtimes. Instead of using an explicit analytical model, NVSHMEM implements this choice as a rule-based decision tree for each collective. Capability checks, datatype and scope constraints, scratch-space availability, and fixed message-size thresholds determine which algorithm is allowed and preferred, with unsupported cases falling back to more general implementations.

\subsection{Support for Multiple Cooperative Thread Arrays (CTAs)}
\label{sec:coll-multi-cta}

The device-side collective APIs in NVSHMEM are threadgroup collectives: they provide thread-, warp-, and block-scoped entry points, but a single collective invocation is executed by \textbf{one participating threadgroup} rather than by multiple CTAs. The absence of public \texttt{\_grid} variants is deliberate. A grid-scoped operation would require synchronization across CTAs inside a running kernel, which is more expensive than ending it and launching a stream-ordered communication kernel. This design is also consistent with NVSHMEM's team model, where each team owns one set of internal collective resources such as \verb|pSync| space, so a team cannot safely host multiple concurrent collective invocations without duplication.

NVSHMEM exposes built-in multi-CTA execution only through the host-side \texttt{\_on\_stream} wrappers, and only for FCollect, AllReduce, and ReduceScatter. On the first multi-CTA call, NVSHMEM creates duplicate teams in a per-team \verb|team_dups[]| array, copies them to the device, and assigns each CTA a separate team and payload slice. Each duplicate team has its own \verb|pSync| region, preserving the rule that one team cannot be used by concurrent collectives. This path is enabled only when NVLS resources are available; otherwise, NVSHMEM uses a single CTA or may fall back to NCCL when NCCL support is enabled. Thus, built-in multi-CTA execution is gated by NVLS availability. Users needing more parallelism typically issue independent block-scoped operations from multiple CTAs or launch a separate communication kernel through an on-stream API.

\subsection{Barrier and Synchronization}

NVSHMEM's synchronization collectives are implemented separately from the data collectives in Table~\ref{tab:collective-algorithms}. In the device code, both \verb|nvshmem_sync| and \verb|nvshmem_barrier| invoke a dissemination-style routine over the team's \verb|pSync| region. At each phase, a PE signals a set of PEs and then waits for the corresponding notifications from another set of partners. The radix \(k\) is chosen from the runtime parameter \verb|barrier_tg_dissem_kval| and clamped by team size. For block-scoped collectives on fully P2P-connected teams, $k$ may be raised to the team size. Under this structure, synchronization requires \(\log_k N\) rounds for a team of \(N\) PEs. 

The difference between \verb|sync| and \verb|barrier| is that \verb|barrier| additionally enforces operation completion and visibility before entering the dissemination rounds via \verb|quiet| or \verb|__threadfence_system| depending on the path, and may also enforce target-side consistency afterward. As a result, both primitives have the same synchronization depth, but \verb|barrier| provides stronger ordering and completion guarantees.

\begin{figure}[t]
  \centering
  \includegraphics[width=\linewidth]{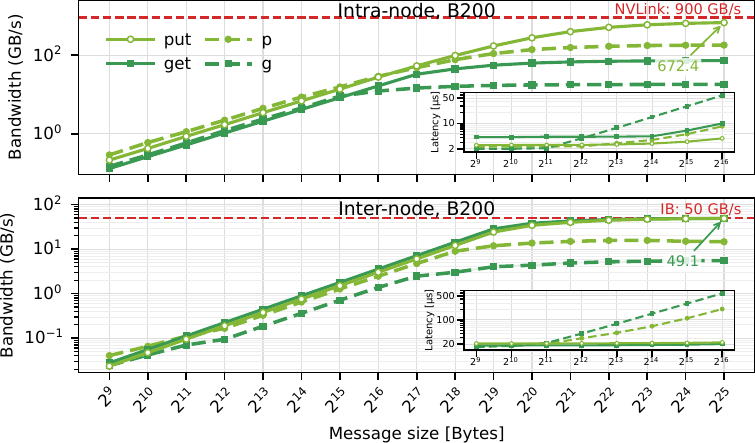}
  \vspace{-2em}
  \caption{B200 one-sided RMA bandwidth from 512~B to 32~MiB. Insets show separately measured one-CTA buffer-completion latency through 64~KiB, using default IBGDA QP settings inter-node. Red dashed lines mark per-direction hardware references: 900~GB/s NVLink and 50~GB/s IB. Colored labels mark peak mean bandwidth, with arrows to that curve's last point. Inter-node bulk \texttt{put}/\texttt{get} nearly overlap. The 512~B starting size excludes empty bulk transfers with 64 CTAs.}
  \label{fig:nvshmem-p2p}
  \vspace{-1.5em}
\end{figure}

\section{Microbenchmarking}

In this section, we present selected microbenchmark results for NVSHMEM’s one-sided RMA and collective operations. As performance evaluation is not the primary focus of this work, we do not attempt a comprehensive study. Rather, these results are intended to highlight a few key performance characteristics of NVSHMEM. Additionally, since NVIDIA already reports official proxy-versus-IBGDA measurements in the GIN paper~\cite{hamidouche2025gin}, we do not collect the same data here.

\subsection{Experimental Setup}

All measurements use DGX B200 nodes on the MARS cluster, each with eight NVIDIA B200 GPUs connected by NVLink-5/NVSwitch with NVLS support. NVLink provides a nominal 1.8~TB/s bidirectional bandwidth per GPU, giving a 900~GB/s per-direction reference. Eight ConnectX-7 NDR 400~Gb/s InfiniBand NICs provide 50~GB/s per NIC and 400~GB/s aggregate per node~\cite{nvidia_dgx_b200}.

The benchmarks use CUDA~13.3, NVSHMEM~3.8.0, and NCCL~2.32.6 with \texttt{nccl-tests}~2.21.1~\cite{nvidia_nccl_tests}. The one-sided RMA bandwidth experiments use NVSHMEM's device point-to-point perftests with 32 CTAs, 256 threads per CTA, 20 warmup and 50 timed iterations intra-node. Inter-node bandwidth runs use IBGDA, 64 CTAs, 1024 threads per CTA, 5 warmup and 20 timed iterations, with \texttt{NVSHMEM\_IBGDA\_NUM\_RC\_PER\_PE=64} and \texttt{NVSHMEM\_IBGDA\_RC\_MAP\_BY=cta}. AllReduce measures \texttt{float} sum. Intra-node NVSHMEM on-stream/device runs use 20/50 warmup and 50/100 timed iterations, respectively. Every plotted point averages eight trials, and shading shows sample standard deviations.

P2P latency is measured separately with the dedicated \texttt{shmem\_\{put,get,p,g\}\_latency} tests. Each uses one CTA: 1024 threads for block-scoped bulk operations and 512 for scalar operations. Runs use 20 warmup and 200 timed iterations, repeated three times per trial. Read tests rotate source addresses across 256 regions to reduce repeated-address caching. Inter-node latency uses IBGDA's default QP settings rather than the bandwidth configuration above.

\subsection{One-sided RMA}

Figure~\ref{fig:nvshmem-p2p} reports device-initiated RMA within and across B200 nodes. Intra-node bulk \texttt{put}/\texttt{get} reach 672/73~GB/s, while scalar \texttt{p}/\texttt{g} reach 181/18~GB/s. Insufficient concurrency, work partitioning, and synchronization can prevent saturation of the 900~GB/s NVLink reference. These are not inherent API ceilings. At 512~B, measured \texttt{put}/\texttt{get}/\texttt{p}/\texttt{g} latency is 2.5/4.2/2.3/2.0~\(\mu\)s. At 64~KiB, scalar \texttt{g} rises to 58.6~\(\mu\)s as each thread serializes multiple element reads.

Across nodes, bulk \texttt{put}/\texttt{get} nearly overlap at 48.9/49.1~GB/s, approaching the 50~GB/s single-NIC reference. Scalar \texttt{p}/\texttt{g} reach 15.7/5.5~GB/s. In the separate latency configuration, 512~B takes 20.4/17.9/16.8/16.0~\(\mu\)s for \texttt{put}/\texttt{get}/\texttt{p}/\texttt{g}. At 64~KiB, bulk operations take 22.1/19.7~\(\mu\)s, versus 205/586~\(\mu\)s for scalar operations, motivating aggregation. The scalar timings cover the entire buffer, not one element. Their growth reflects multiple element-wise operations per thread and return-value dependencies for reads.

\begin{figure}[t]
  \centering
  \includegraphics[width=\linewidth]{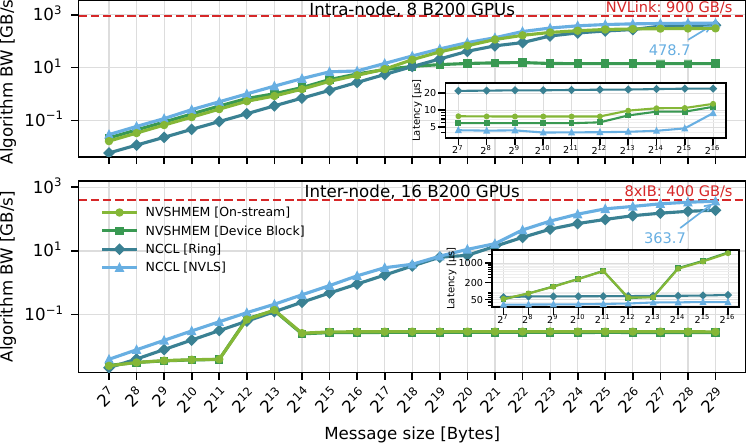}
  \vspace{-1em}
  \caption{B200 AllReduce algorithm bandwidth ($M/t$). Insets show measured collective latency through 64~KiB. Red dashed lines mark hardware references: 900~GB/s per-direction NVLink per GPU and $8\times50=400$~GB/s IB per node, since the collective can use all eight NICs concurrently. Algorithm bandwidth is not individual-link throughput. Colored labels indicate the highest mean bandwidth in each panel, with arrows pointing to that curve's last point.}
  \label{fig:nvshmem-allreduce}
  \vspace{-1.5em}
\end{figure}

\subsection{Collective: AllReduce}

For collective benchmarking, we consider only AllReduce, since it is one of the most performance-critical collectives in many applications~\cite{tang2024recent, chrapek2023hear, adam2023xccl, xiong2024revisiting} and is the operation for which both NVSHMEM and NCCL provide highly optimized implementations. Figure~\ref{fig:nvshmem-allreduce} compares their intra-node and inter-node performance. \texttt{NVSHMEM [On-stream]} uses the host-side \verb|_on_stream| path, enabling multi-CTA execution and NVLS-based algorithms for the intra-node case, while \texttt{NVSHMEM [Device Block]} restricts execution to a fully device-initiated single-CTA path. \texttt{NCCL [Ring]} denotes the standard Ring AllReduce, whereas \texttt{NCCL [NVLS]} uses symmetric kernels such as \verb|RSxLDMC_AGxSTMC| intra-node and the \verb|NVLS Tree| algorithm inter-node for messages of at least 4~MiB, with Tree used for smaller messages.

Intra-node, NVSHMEM on-stream reaches 303~GB/s over the plotted sizes below 1~GiB, versus 15.4~GB/s for the single-CTA device path, demonstrating the benefit of multi-CTA execution. NCCL ring and its NVLS-enabled path reach 398/479~GB/s. At 64~KiB, measured device/on-stream latency is 11.5/12.9~\(\mu\)s, versus 24.0/8.8~\(\mu\)s for NCCL ring/NVLS.

Across nodes, NVSHMEM's multi-CTA NVLS path is unavailable for the full team. With the default IBRC host proxy and 512~KiB reduction scratch, both paths peak below 0.14~GB/s, versus NCCL ring/NVLS Tree at 191/364~GB/s. At 128~B, NVSHMEM takes 51--52~\(\mu\)s, slightly below ring's 60~\(\mu\)s but above NCCL's NVLS-enabled configuration at 32~\(\mu\)s. At 64~KiB, NVSHMEM rises to 2.3--2.4~ms versus NCCL's 72/40~\(\mu\)s. This gap reflects the measured segmented, single-CTA collective configuration, not the tuned IBGDA P2P path above or a general limit on NVSHMEM communication.

\section{Case Study: DeepEP}

In this case study, we examine how NVSHMEM is used in real applications and communication libraries through DeepEP~\cite{deepep2025}, an open-source library released by DeepSeek for Mixture-of-Experts (MoE) workloads under expert parallelism (EP). In EP, the experts of each MoE layer are sharded across GPUs. For every layer, tokens must be \textbf{dispatched} to the experts selected by the router, and after computation, the outputs are \textbf{combined} back to the GPUs that own the original tokens. Both phases are sparse all-to-all exchanges, where the communication volume between pairs of ranks is data-dependent. This makes them difficult to implement efficiently with off-the-shelf collective libraries. DeepEP addresses this challenge with custom dispatch and combine kernels, using NVSHMEM as the substrate for cross-node RDMA.

DeepEP exposes both a \textbf{high-throughput (HT)} path for training and a \textbf{low-latency (LL)} path for inference. In the rest of this section, we explain these two paths in more detail. Our analysis focuses on DeepEP V1, since the newer V2 backend is based on NCCL GIN and is beyond the scope of this work.

\subsection{High-Throughput (HT) Kernels}

\begin{figure}[t]
  \centering
  \includegraphics[width=\linewidth]{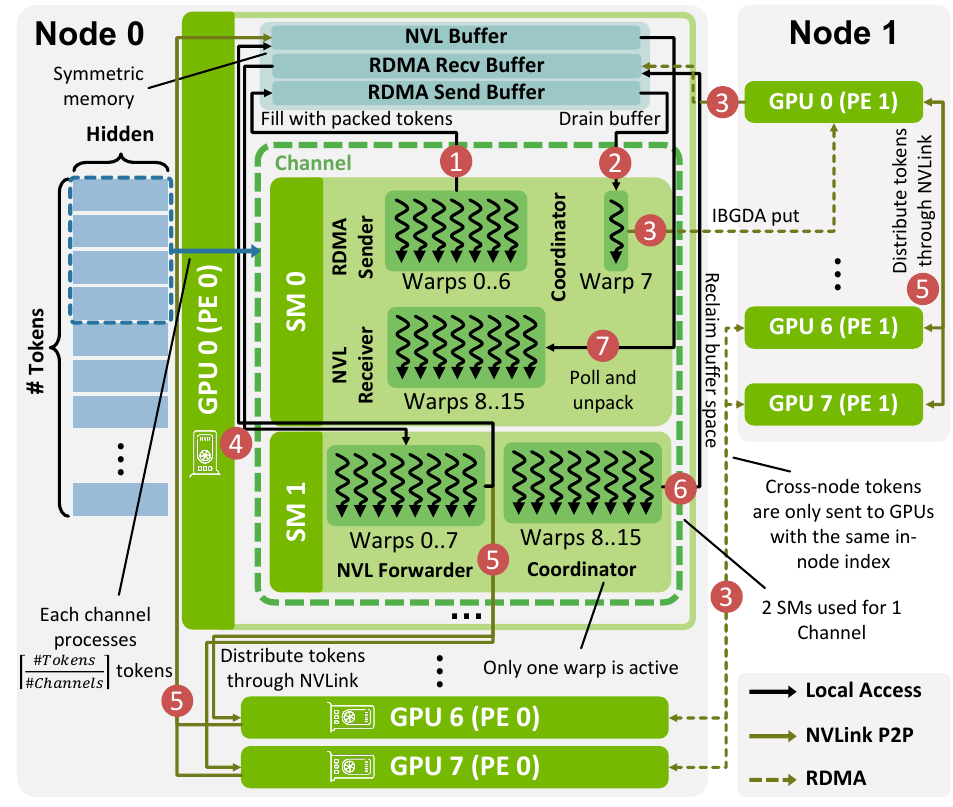}
  \vspace{-2em}
  \cprotect\caption{Illustration of the HT dispatch algorithm for a two-node configuration. The final step in which the NVLink receiver warps store the tokens into the output is omitted. The numbered markers \circled{1}--\circled{7} show the main steps taken.}
  \label{fig:deepep-ht-dispatch}
  \vspace{-1.5em}
\end{figure}

DeepEP’s high-throughput path targets training, where large token counts make bandwidth more important than per-token latency. Its key idea is a two-stage pipeline: tokens are first sent across nodes by RDMA only between GPUs with the same local index, and are then redistributed within the destination node over NVLink to the GPUs that host the selected experts. The main components and steps of the HT dispatch algorithm are illustrated in Figure~\ref{fig:deepep-ht-dispatch}.

A key implementation detail is that the HT path uses \emph{eight} parallel NVSHMEM world teams, one for each GPU slot in a node. Each world contains one PE per node, so cross-node RDMA happens only between GPUs with the same local index. However, it also assumes exactly eight P2P-accessible GPUs per node, so the design is less portable to other systems.

The communication itself is divided into two phases. The first phase, \texttt{notify\_dispatch}, exchanges metadata rather than payloads. It gathers per-rank and per-expert token counts, computes prefix matrices, and determines how each channel will partition its share of the outgoing traffic. The second phase, \texttt{dispatch} (and symmetrically \texttt{combine}), moves the actual payloads. Each dispatched token is packed into a special wire format containing information such as the hidden state, optional scaling metadata, and routing information.

The dispatch kernel is organized into logical \textbf{channels}, each spanning a pair of Streaming Multiprocessors (SMs) and handling a contiguous slice of the input tokens. Within each SM, DeepEP uses \textbf{warp specialization} across its 16 warps. On the RDMA-side SM, 7 warps act as senders, 1 warp serves as a sender coordinator, and the remaining 8 act as NVLink receivers, one for each local GPU slot. On the paired SM, 8 warps serve as RDMA-to-NVLink forwarders, one per destination GPU slot, while the remaining 8 are forwarder coordinators, although only one of them is active in practice.

During dispatch, the sender warps place tokens into per-peer RDMA ring buffers on the symmetric NVSHMEM heap, and the sender-coordinator warp periodically batches these writes into larger RDMA transfers through \texttt{nvshmemi\_ibgda\_put\_nbi\_warp}, updating the remote tail with \texttt{nvshmemi\_ibgda\_amo\_nonfetch\_add}. On the receive side, the 8 forwarder warps poll for arrivals, decode token metadata, and copy tokens into intra-node NVLink ring buffers. A single forwarder-coordinator warp tracks the forwarders’ progress and returns RDMA credits, while the 8 NVLink-receiver warps place the tokens into the output tensor.

The \texttt{combine} kernel follows the same structure in reverse. Expert outputs are first moved locally over NVLink into RDMA buffers, then sent across nodes, and finally reduced into the destination tensor on the token’s original GPU. The warp roles differ slightly, but the overall design is the same.

A key point is that DeepEP does not express most work as NVSHMEM collectives or generic RMA calls. Instead, it uses NVSHMEM only at critical points in the inter-node RDMA path: metadata exchange in \texttt{notify\_dispatch}, chunked RDMA puts from sender coordinators, and atomic credit updates for remote head or tail counters. Thus, NVSHMEM acts as the cross-node communication substrate, while DeepEP builds its own multi-stage transport pipeline on top.

\subsection{Low-Latency (LL) Kernels}

DeepEP's low-latency path targets inference, where batch sizes are small and end-to-end layer latency matters more than peak bandwidth. Accordingly, it removes the intra-node NVLink forwarding stage used by the high-throughput path and relies on RDMA for inter-node delivery.

Additionally, instead of partitioning the cluster into eight independent NVSHMEM world teams, the low-latency path uses a single global NVSHMEM world team and overlays a strided team for the GPUs with the same slot across nodes.

The kernel structure is likewise much simpler. There are no logical channels and warp specialization. Instead, a single kernel grid handles the whole dispatch, and SMs are partitioned by \emph{local experts} rather than by token range. Within each block, warps are partitioned into fixed-size groups, and each warp group is responsible for one expert.

Both dispatch and combine follow this simple structure. In the send phase, each warp iterates over tokens and handles one top-\(k\) destination per warp. After optional FP8 conversion, lane 0 reserves a slot for the destination expert and computes the target address in the remote receive buffer. If the destination is on another node, the warp sends the packed message with \verb|nvshmemi_ibgda_put_nbi_warp|; if it is on the same node and P2P-accessible, the warp instead performs a direct local copy into the mapped receive buffer. After all messages for an expert have been sent, one atomic update publishes the final token count to the destination.

In the receive phase, the corresponding warp group waits until the count becomes nonzero, reads the number of received tokens, and reserves space in the output buffer for that expert. It then copies the messages from the receive slots into the output tensor, unpacking source indices and optional FP8 scales. Overall, the critical-path NVSHMEM activity is minimal: the LL path uses an IBGDA put for payload transfer and an atomic update for the final count notification.

\subsection{Performance}

NVIDIA has already published a direct comparison of NCCL GIN and NVSHMEM on DeepEP, and since the purpose of this paper is to explain NVSHMEM rather than to add another benchmarking study, we do not repeat those experiments here. Instead, we briefly summarize the GIN paper’s results. There, DeepEP was integrated with NCCL GIN while the original NVSHMEM-based version was retained as the baseline, and both the high-throughput (HT) and low-latency (LL) dispatch and combine kernels were evaluated. The main takeaway is that NCCL GIN matches NVSHMEM closely, typically within about 1--2\%, while offering the same style of device-initiated communication inside NCCL’s runtime.

\section{Related Work \& Discussion}

The closest work to ours is that of Hu et al.~\cite{hu2026demystifyingnccl}, which presents a source-level analysis of NCCL. Among prior works on NVSHMEM itself, the most directly related is that of Langer et al.~\cite{langer2022dynamic}, which explains dynamic symmetric heap. While informative, that study is limited to the memory subsystem, whereas our work analyzes the library more broadly.

A separate line of work studies NVSHMEM from the perspective of applications and runtimes. Hsu et al.~\cite{hsu2020assessment} evaluate NVSHMEM in terms of usability and functionality, while later works use it in systems such as GROMACS and CharminG to improve scaling~\cite{doijade2025redesigning, choi2021charming}. More broadly, related work on GPU-centric programming models, such as Hamidouche et al.~\cite{hamidouche2020gpuinitiated} and Unat et al.~\cite{unat2024landscape}, examines the design space around GPU-resident communication mechanisms. In contrast, our contribution is not to propose a new runtime or survey the entire space, but to provide a detailed study of NVSHMEM.

One clear takeaway from our analysis is that limited multi-CTA support remains a weakness of NVSHMEM’s built-in collective implementation, as discussed in Section~\ref{sec:coll-multi-cta}. Meanwhile, NCCL is narrowing the gap through its device API. However, the GIN paper also shows that NVSHMEM still provides a meaningful advantage as a lower-level one-sided RMA substrate~\cite{hamidouche2025gin}. NVSHMEM, therefore, remains relevant for applications that need fine-grained one-sided communication and direct device-side control.

\section{Conclusion}

NVSHMEM occupies a distinct point in the GPU communication design space by enabling GPU-initiated, one-sided communication through a PGAS model. Our source-level analysis shows how this model is implemented through symmetric memory, multiple transport paths, and a layered runtime that connects device-side operations to underlying network mechanisms. We also show that NVSHMEM’s main limitations today come from its execution model and collective implementations, which often do not fully exploit GPU parallelism. The DeepEP case study highlights NVSHMEM’s practical value in modern sparse deep learning workloads. Overall, NVSHMEM has made an important contribution by exposing a flexible, device-driven substrate for fine-grained data movement, and its design continues to drive the evolution of modern GPU communication libraries.

\section{Acknowledgment}
The research was conducted as part of the FastTrackAI project at the Singapore-ETH Centre, which was established collaboratively between ETH Zurich and the National Research Foundation, Singapore. This research is supported by the National Research Foundation, Singapore (NRF), and the Ministry of Digital Development and Information (MDDI) under the AI Visiting Professorship (Award No. AIVP-2025-005). This work also received funding from the European Research Council (Project PSAP, No. 101002047). The authors used GPT-5.5 to assist with light editing and proofreading. All content and ideas remain the original work of the authors.

\bibliographystyle{ieeetr}
\bibliography{references}

\appendix

\begin{table*}[t]
    \centering
    \caption{Breakdown of NVSHMEM API groups by availability and functionality in the source code.}
    \label{tab:api-groups}
    \renewcommand{\arraystretch}{1.5}
    \resizebox{1\textwidth}{!}{%
    \begin{tabular}{>{\bfseries}p{0.15\textwidth}p{0.18\textwidth}p{0.63\textwidth}}
    \toprule
    \textbf{API group} & \textbf{Host/Device Availability} & \textbf{Description} \\[2pt] \hline

    \rowcolor{lightgrayrow}
    Setup / exit & Mixed & Initialize and finalize the library, bootstrap jobs, and handle exceptional termination. \\

    Memory management & Host-side & Allocate, free, and align symmetric objects in the symmetric heap, provide registration functionality for selected local buffers. \\

    \rowcolor{lightgrayrow}
    Team management & Mixed, mostly host-side & Define communication subgroups, query team-relative PE identities, translate PE numbers across teams, and create or destroy user-defined teams. \\

    One-sided RMA & Both & Provide one-sided data movement through \texttt{put}, \texttt{get}, scalar \texttt{p}/\texttt{g}, etc. \\

    \rowcolor{lightgrayrow}
    Atomics & Both & Provide remote atomic read-modify-write operations such as fetch, add, compare-and-swap, and bitwise updates on symmetric data. \\

    Memory ordering & Both & Provide ordering and completion primitives such as \texttt{fence} and \texttt{quiet}. \\

    \rowcolor{lightgrayrow}
    Synchronization & Mixed, mostly device-side & Provide value-based synchronization through \texttt{wait} and \texttt{test} operations. \\
    
    Collectives & Both & Provide coordinated multi-PE operations such as AllReduce, and Broadcast. \\

    \hline
    \end{tabular}%
    }
\end{table*}

\end{document}